\documentclass[a4paper,11pt]{article}
\usepackage[utf8]{inputenc}
\usepackage[english]{babel}
\usepackage{braket}
\usepackage{xspace}
\usepackage{bbm}
\usepackage{mathdots}
\usepackage{stackrel}
\usepackage[dvipsnames]{xcolor}
\usepackage{appendix}
\usepackage{hyperref}
\hypersetup{
 colorlinks=true,
 linkcolor=blue,
 anchorcolor = blue,
 citecolor = blue,
 filecolor = blue,
 urlcolor = blue
}
\usepackage{mathtools}
\usepackage{bbm}
\usepackage{comment}
\usepackage{subfigure}
\usepackage[T1]{fontenc}               
\usepackage[left=3cm,
            right=2.5cm,
            top=2.5cm,
            bottom=3cm
            ]{geometry}                
\usepackage{graphicx}
\usepackage{mathrsfs}
\usepackage{appendix}
\usepackage{fancyhdr}
\usepackage{amsmath,                   
            amssymb,                   
            amsthm}                    

\usepackage{setspace}
\usepackage{cite}
\usepackage{cancel}
\usepackage[margin=30pt, bf, font=small, center, justification=justified]{caption}[2004/07/16]

\usepackage{setspace}

\title{\bf More on symmetry resolved operator entanglement 
}

\author{Sara Murciano$^{1,2}$ J\'er$\hat{\mathrm{o}}$me Dubail$^{3}$, Pasquale Calabrese$^{4,5}$}

\date{}

\begin{document} 

\maketitle
{\small
\vspace{-5mm}  \ \\
$^{1}$	Walter Burke Institute for Theoretical Physics, Caltech, Pasadena, CA 91125, USA\\
\medskip
$^{2}$	Department of Physics and IQIM, Caltech, Pasadena, CA 91125, USA\\[-0.1cm]
\medskip
$^{3}$ Universit\'{e} de Lorraine, CNRS, LPCT, F-54000 Nancy, France\\[-0.1cm]
\medskip
$^{4}$  SISSA and INFN Sezione di Trieste, via Bonomea 265, 34136 Trieste, Italy\\
\medskip
$^{5}$ International Centre for Theoretical Physics (ICTP), Strada Costiera 11, 34151 Trieste, Italy\\
\medskip
}

\begin{center}
\date{\today}    
\end{center}

\maketitle

\begin{abstract}
The `operator entanglement' of a quantum operator $O$ is a useful indicator of its complexity, and, in one-dimension, of its approximability by matrix product operators. Here we focus on spin chains with a global $U(1)$ conservation law, and on operators $O$ with a well-defined $U(1)$ charge, for which it is possible to resolve the operator entanglement of $O$ according to the $U(1)$ symmetry. 
We employ the notion of symmetry resolved operator entanglement (SROE) introduced in [\href{https://doi.org/10.1103/PRXQuantum.4.010318}{Rath {\it et al.}, PRX Quantum {\bf 4}, 010318 (2023)}] and extend the results of the latter paper in several directions. 
Using a combination of conformal field theory and of exact analytical and numerical calculations in critical free fermionic chains, we study the SROE of the thermal density matrix $\rho_\beta = e^{- \beta H}$ and of charged local operators evolving in Heisenberg picture $O = e^{i t H} O e^{-i t H}$. 
Our main results are: 
i) the SROE of $\rho_\beta$ obeys the operator area law;
ii) for free fermions, local operators in Heisenberg picture can have a SROE that grows logarithmically in time or saturates to a constant value;
iii) there is equipartition of the entanglement among all the charge sectors except for a pair of fermionic creation and annihilation operators. 
\end{abstract}

\tableofcontents

\section{Introduction}

Quantum entanglement plays a pivotal role to understand emergent phenomena in quantum many-body physics and numerical methods. In this context, the entanglement entropy has received significant attention and has become the most popular measure of bipartite entanglement in quantum systems. It is relevant in various contexts, ranging from high-energy physics~\cite{RT, dong, maldacena, Raamsdonk} to condensed matter theory~\cite{intro1,intro2,eisert-2010,intro3}, when studying extended systems like quantum field theories (QFTs) and lattice models. For example, it can be useful to detect and describe phase transitions, even when a conventional order parameter is unavailable. In fact, its behaviour as a function of the subsystem size allows us to discern if the system is in a gapped or gapless phase and what are the universal features of critical systems. It turns out that in the former case, the entanglement follows the area law~\cite{h-06}, i.e. it is proportional to the size of the border of the subsystem, in contrast to the thermal entropy which is characterised by a volume law. However, for the ground state of gapless local Hamiltonians in one dimension (1D), the area law is corrected by a logarithmic term~\cite{hlw-94,cw-94,gv-03,cc-04,cc-09}. 

One practical implication is that the presence of entanglement makes it difficult to simulate quantum many-body systems on a classical computer. For example, efficient simulations based on Matrix Product States (MPS) work well in non critical 1D systems~\cite{orus-19,vidal-04,vmc-08,dmrg,tdmrg}, but they are less efficient when we approach a quantum critical point. The counterpart to MPS for approximating operators are Matrix Product Operators (MPOs), which are tensor network representations of operators, and it is natural to ask whether a quantity similar to the entanglement entropy exists to accurately capture the validity of this approximation~\cite{Prosen2007,Pizorn2009,dubail,Zhou2017,noh2020efficient,rakovszky2022dissipation,Alba2022rise,pablo2023}. The answer lies in the concept of operator entanglement (OE), which quantifies the entanglement between quantum operators acting on different parts of a quantum system~\cite{Prosen2007,Pizorn2009,zanardi2000entangling,zanardi2001entangling,dubail,Zhou2017,jonaynahum,alba1,bruno1,bruno2,alba2,pavel,tagliacozzo,fleischhauer,Styliaris_2021,barch2023scrambling,gillesOE}.
The results about the OE depends on the specific operator and on the framework in which it is employed. 
In this paper we continue the analysis initiated in Ref.~\cite{rath-22} of the symmetry resolved operator entanglement (SROE), which extends the notion of OE to consider the entanglement properties of operators with respect to specific symmetries of the quantum system. 
The interplay between the entanglement of a state and symmetries has been intensively studied in the last years through the symmetry-resolved entropies~\cite{lr-14,sierra,SG17}, both theoretically~\cite{mdc-19,trac-20,mdgc-20,bpc-21,nonabelian-21,capizzi,dhc-20,znm-20,regnault} and experimentally~\cite{fis,vitale,neven,rath-22}, for several entanglement measures~\cite{Capizzi-Cal-21,goldstein1,mbc-21,fidelity,dge-23}. 
The only symmetry resolution for operators studied so far is the reduced density matrix after a quantum quench from a product state \cite{rath-22} (see also for related quantities in open systems \cite{pablo2023,alba1}). As in the non-resolved case, it 
exhibits an entanglement barrier~\cite{dubail,Wang2019barrier,Reid2021Entanglement}: it grows linearly in time as
entanglement builds up between the local degrees of freedom, it then reaches a maximum, and ultimately decays to a small finite value as the reduced density matrix converges to a simple stationary
state. 

Before moving to the organisation of the paper, we introduce the main concepts: the OE, and how to symmetry resolve it.

\subsection{Definition of Operator Entanglement}
The entanglement properties between a region $A$ and its complementary $B$ are usually defined starting from a state $\left| \psi \right>$. Nevertheless, one can also study the entanglement properties of an operator $O$ by vectorizing it. Namely, if the operator $O$ lives in $\mathrm{End}(\mathcal{H}_A)\otimes \mathrm{End}(\mathcal{H}_B)$, we can view it equivalently as a vector in $(\mathcal{H}_A\otimes\mathcal{H}_B) \otimes (\mathcal{\bar{H}}_A\otimes\mathcal{\bar{H}}_B)$, where $\mathcal{H}_{A,B}$ denotes the Hilbert space in part A or B, and $\mathcal{\bar{H}}_{A,B}$ its dual. 
To work with a properly normalised state after the vectorization, we divide the operator $O$ by a normalisation factor $\sqrt{\mathrm{Tr}(O^{\dagger}O)}$. Operationally, one can pick an orthonormal basis $\{\ket{i} \}$ for $(\mathcal{H}_A\otimes\mathcal{H}_B)$ and the corresponding basis $\{\ket{j} \}$ for its dual $(\mathcal{\bar{H}}_A\otimes\mathcal{\bar{H}}_B)$, and write $O=\sum_{ij}O_{ij}\ket{i} \bra{j}$, where $O_{ij}=\bra{i}O\ket{j}$. Then the normalised operator-state is obtained as~\cite{choi1,choi2}
\begin{equation}\label{eq:vectorisation}
\ket{O}=\frac{1}{\sqrt{\mathrm{Tr}(O^{\dagger}O)}}\sum_{ij}O_{ij}\ket{i} \ket{j}.
\end{equation}
Importantly, $O$ admits a Schmidt decomposition,
\begin{equation}\label{eq:schmidt}
\frac{O}{\sqrt{\mathrm{Tr}(O^{\dagger}O)}}=\sum_{i=1}^r \lambda_i O_{A,i} \otimes O_{B,i} ,
\end{equation} 
where $r$ is the operator Schmidt rank, and the $\lambda_i$ are real positive coefficients that satisfy $\sum_{i=1}^r \lambda_i^2 = 1$. The operators $O_{A,i} \in \mathrm{End}(\mathcal{H}_A)$ (same for $O_{B,i}$) obey the orthonormality condition $\mathrm{Tr}[O_{A,i}^{\dagger},O_{A,j}]=\delta_{ij}$. This can be seen by performing the ordinary Schmidt decomposition of the pure state $\ket{O}$ and eventually reverting the vectorization to get back to the space of operators.

The OE is defined as follows. From the vectorization $\ket{O}$, we can build the super-reduced-density-matrix $\mathrm{Tr}_{B\otimes B}(\ket{O}\bra{O})$ ---where `super' refers to the operators in the space of operators--- which is a super-operator acting on operators on $\mathcal{H}_A$. Then the OE is the R\'enyi entropy of that super-reduced-density-matrix,
\begin{equation}
S_n(O)=\frac{1}{1-n}\log\mathrm{Tr}[ (\mathrm{Tr}_{B\otimes B}(\ket{O}\bra{O}) )^n]. 
\end{equation}
Alternatively, the OE is given in terms of the Schmidt values $\lambda_i$ in Eq.~(\ref{eq:schmidt}) as
\begin{equation}\label{eq:1def}
S_n(O)=\frac{1}{1-n}\log \sum_{i=1}^r(\lambda_i^2)^n.
\end{equation}
As usual, the limit $n\to 1$ produces the von Neumann OE
\begin{equation}\label{eq:1def}
S(O)=-\sum_{i=1}^r\lambda^2_i\log \lambda^2_i.
\end{equation}

\subsection{$U(1)$ charge and symmetry resolution of OE}
We now assume that there is a charge operator $Q$ acting on the full Hilbert space $\mathcal{H} = \mathcal{H}_A \otimes \mathcal{H}_B$ which generates a $U(1)$ symmetry, and which is a sum of the two charge operators acting on subsystems $A$ and $B$, i.e. $Q=Q_{A}\otimes \mathbbm{1}_B+\mathbbm{1}_A\otimes Q_{B}$. A natural question is how to define a symmetry resolution of the OE of an operator $O$ that possesses a fixed charge $q_O$, i.e. $[Q,O]=q_O Q$. The answer is based on the symmetry resolution for $\ket{O}$, as we are going to review. This problem has been already addressed in~\cite{rath-22,Alba2022rise,pablo2023} and we report here the main definitions we will use in this manuscript.

From the commutation relation between $O$ and $Q$, the terms in the Schmidt decomposition~\eqref{eq:schmidt} can be reorganised according to their charge $q$ as~\cite{rath-22,dge-23}
\begin{equation}
    \label{eq:OABgeneral}
   \frac{O}{\sqrt{\mathrm{Tr}(O^{\dagger}O)}}= \sum_q \sum_j \lambda^{(q )}_j O_{A,j}^{(q)} \otimes O_{B,j}^{(q_O-q)},
\end{equation}
where 
\begin{equation}\label{eq:qdefinition}
    \Big[Q_A , O_{A,j}^{(q)}\Big] = q \, O_{A,j}^{(q)}, \quad  \Big[Q_B , O_{B,j}^{(q_O-q)}\Big] = (q_O-q) \, O_{B,j}^{(q_O-q)},
\end{equation}
such that \mbox{$ [Q, O_{A,j}^{(q)} \otimes O_{B,j}^{(q_O-q)} ] = q_O  O_{A,j}^{(q)} \otimes O_{B,j}^{(q_O-q)}$}.
The charge $q$ that appears in these equations can be introduced as the eigenvalue of a charge `superoperator' $\mathcal{Q}$ living in the Hilbert space $\mathrm{End}(\mathcal{H}) \otimes \mathrm{End}(\bar{\mathcal{H}})$ 
\begin{equation}\label{eq:charge}
\mathcal{Q}=Q\otimes \mathbbm{1}-\mathbbm{1}\otimes Q^T.
\end{equation}
Such superoperator satisfies the commutation relation 
\begin{equation}
[\ket{O}\bra{O},\mathcal{Q}]=0,
\end{equation}
where we have used the vectorization introduced in Eq.~\eqref{eq:vectorisation}. Using the local structure of $Q$ in $A\cup B$, we can write
\begin{equation}\label{eq:chargeA}
\mathcal{Q}=\mathcal{Q}_A \otimes \mathbbm{1}_B+ \mathbbm{1}_A\otimes\mathcal{Q}_B,\quad \mathcal{Q}_A=Q_A\otimes \mathbbm{1}_A-\mathbbm{1}_A\otimes Q_A,
\end{equation}
and the following commutation relation holds
\begin{equation}\label{eq:comm1}
[\mathrm{Tr}_{B\otimes B}(\ket{O}\bra{O}),\mathcal{Q}_A]=0.
\end{equation}
We can exploit the result above such that $ \mathrm{Tr}_{B\otimes B}(\ket{O}\bra{O})$ can be decomposed as
\begin{equation}\label{eq:dec1}
  \mathrm{Tr}_{B\otimes B}(\ket{O}\bra{O})=\bigoplus_q   p(q) \mathrm{Tr}_{B\otimes B}(\ket{O}\bra{O})(q), \quad p(q)\equiv\mathrm{Tr}\left[\Pi_q  \mathrm{Tr}_{B\otimes B}(\ket{O}\bra{O})\right],
\end{equation}
where $\Pi_q$ is a projector onto the eigenspace of $\mathcal{Q}_A$ with fixed $q$ and $\mathrm{Tr}_{B\otimes B}(\ket{O}\bra{O})(q)$ denotes the (normalised) reduced density matrix built from the vector $\ket{O}$ and restricted to the charge $q$. The partition function projected in a given charge sector reads
\begin{equation}\label{eq:Zcal}
    \mathcal{Z}^{(n)}_{q}(O)\equiv\frac{\mathrm{Tr}[\Pi_q \left(\mathrm{Tr}_{B\otimes B}(\ket{O}\bra{O})\right)^{n}]}{\mathrm{Tr}(O^{\dagger}O)^{n}},
\end{equation}
and the SROE is given by
\begin{equation}\label{eq:FTOE}
\begin{split} S_{q}^{(n)}(O)&=\frac{1}{1-n}\log \frac{\mathcal{Z}^{(n)}_q(O)}{[\mathcal{Z}^{(1)}_{q}(O)]^n},\qquad S_{q}(O)=\displaystyle \lim_{n\to 1} S_{q}^{(n)}(O).
\end{split}
\end{equation}
According to Eq.~\eqref{eq:dec1}, the total von Neumann OE associated to $O$ splits into 
\begin{equation}
\label{eq:SvN}
S(O)=\displaystyle \sum_q p(q) S_q(O)- \displaystyle \sum_q p(q)  \log p(q) .
\end{equation}
As it was shown in detail in~\cite{rath-22},  using the uniqueness of the Schmidt coefficients, the set of all (non-zero) values $\{\lambda_j^{(q)}\}$ altogether must be the same as the set of values $\{\lambda_i\}$ from Eq.~\eqref{eq:schmidt}. Therefore, all quantities defined above can be defined in terms of $\{\lambda_j^{(q)}\}$ as follows
\begin{equation}\label{eq:pqsroe}
p(q) = \sum_{j} (\lambda_j^{(q)})^2,\qquad   S_q^{(n)} (O) \,  = \, \frac{1}{1-n}  \log \left( \sum_j \left(\frac{ (\lambda_j^{(q)})^2}{p(q)} \right)^{\!\!n\,} \right).
\end{equation}

\subsection{Organization of the paper}

Our goal is to extend our previous study on SROE \cite{rath-22}. The manuscript is organised as follows. In Section~\ref{sec:techniques} we review the known analytical techniques for computing the (non-resolved) OE in CFT and in free fermion spin chains, and we extend these to include the symmetry resolution. In Section~\ref{sec:thermal} we apply these techniques to the 
case of the density matrix of a critical 1D system at thermal equilibrium. In Section~\ref{sec:heisenberg_picture} we study the SROE of local operators evolving in Heisenberg picture for free fermion chains. We draw our conclusions in Section~\ref{sec:conclusion}.

%
%

\section{Techniques to compute the SROE}
\label{sec:techniques}

Before reviewing the technical tools necessary to evaluate the SROE, both in field theory and in free fermionic lattice models, we summarise how to tackle the problem of symmetry resolution of a $U(1)$ symmetric state following Refs.~\cite{SG17,mdgc-20}.

\subsection{$U(1)$ symmetry resolution}
In this section, we explain how to compute the symmetry resolved entanglement entropy in a given charge sector for a state $\ket{\psi}$, denoted by $S_{q}^{(n)}(\ket{\psi})$.

Let us consider a system with an internal $U(1)$ symmetry and its bipartition into two subsystems, $A$ and $B$.  
Tracing out the degrees of freedom of $B$, we obtain the reduced density matrix (RDM) of $A$, $\rho_A=\mathrm{Tr}_B \ket{\psi}\bra{\psi}$. A measure of the entanglement between $A$ and its complementary part is provided by the R\'enyi entropies, defined as
\begin{equation}
S^{(n)}(\ket{\psi})=\frac{1}{1-n} \ln\mathrm{Tr} \rho_A^{n},
\end{equation}
and the limit $n \to 1$ gives the von Neumann entropy as usual. When $\ket{\psi}$ is an eigenstate of the hermitian charge operator $Q$, by taking the partial trace of the commutator $[ \ket{\psi}\bra{\psi},Q]=0$ over $B$ one finds that $[\rho_A,Q_A] = 0$, using $Q=Q_{A}\otimes \mathbbm{1}_B+\mathbbm{1}_A\otimes Q_{B}$. This means that the reduced density matrix $\rho_A$ has a block-diagonal structure where each block corresponds to an eigenvalue $q'$ of $Q_A$,
\begin{equation}
\label{eq:sum}
\rho_A=\oplus_{q'} p_{A}(q')\rho_{A}(q'),
\end{equation}
where $p_{A}(q')$ is the probability of finding $q'$ in a measurement of $Q_A$ in the RDM $\rho_A$, i.e. $p_A(q')= \mathrm{Tr} \left(\Pi_{q'} \rho_A\right)$. 
Within this convention, the density matrices $\rho_A (q')$ of different blocks are normalised as ${\rm tr}\rho_A (q')=1$.  
Thus, from the normalised $\rho_A(q')$, we can define the {\it symmetry resolved R\'enyi entropies} as
\begin{equation}
\label{eq:RSREE}
S_{q'}^{(n)}(\ket{\psi}) \equiv \dfrac{1}{1-n}\ln \mathrm{Tr} \rho^n_A(q')\quad S_{q'}(\ket{\psi})=\lim_{n\to 1} S_{q'}^{(n)}(\ket{\psi}).
\end{equation}
The total von Neumann entanglement entropy associated to $\rho_A$ in Eq. (\ref{eq:sum}) admits a decomposition as in Eq.~\eqref{eq:SvN},
\begin{equation}
\label{eq:SvN2}
S(\ket{\psi})=\displaystyle \sum_{q'} p(q') S_{q'}(\ket{\psi})- \displaystyle \sum_{q'} p(q') \ln p(q').
\end{equation}
The two terms are known as `configurational entanglement entropy' and `fluctuation entanglement entropy' (or `number entanglement entropy') respectively~\cite{fis}. 
The configurational entropy is also related to the operationally accessible entanglement entropy of Refs.~\cite{wv-03,bhd-18,bcd-19}, 
while the number entropy is the subject of a substantial recent activity~\cite{fis,kusf-20,kusf-20b,riccarda,ms-20}. 

The calculation of the symmetry resolved entropies by the definition (\ref{eq:RSREE}) is a difficult task, especially for an analytic derivation.
As proposed in Ref.~\cite{SG17}, it is convenient to focus on the charged moments of $\rho_A$, $\mathrm{Tr}\left(\rho_A^ne^{iQ_A \alpha}\right)$~\cite{Belin-Myers-13-HolChargedEnt,cms-13,cnn-16,d-16}. Their Fourier transforms with respect to $\alpha$ are the moments of the RDM restricted to the sector of fixed charge $q'$~\cite{SG17}, i.e.
\begin{equation}
\label{eq:defF}
\mathcal{Z}^{(n)}_{q'}(\ket{\psi})\equiv \mathrm{Tr} (\Pi_{q'}\,\rho^n_A)=\displaystyle \int_{-\pi}^{\pi}\dfrac{d\alpha}{2\pi}e^{-iq'\alpha}\mathrm{Tr}\left(\rho_A^ne^{iQ_A \alpha}\right).
\end{equation}
Finally the symmetry resolved entropies are obtained as
\begin{equation}
\label{eq:SREE1}
S^{(n)}_{q'}(\ket{\psi})=\dfrac{1}{1-n}\ln \left[ \dfrac{\mathcal{Z}^{(n)}_{q'}(\ket{\psi})}{\left(\mathcal{Z}^{(1)}_{q'}(\ket{\psi})\right)^n}\right].
\end{equation}
We can apply the same machinery to compute the symmetry resolution of the OE in the charge sectors of the operator in Eq.~\eqref{eq:chargeA} starting from the charged moments of the super-density-matrix built from $\ket{O}$. They are defined as
\begin{equation}\label{eq:charged_moments}
    Z_n(\alpha)\equiv \mathrm{Tr} [\left(\mathrm{Tr}_{B\otimes B}(\ket{O}\bra{O})\right)^ne^{i\alpha\mathcal{Q}_A}],
\end{equation}
and their Fourier transform gives
\begin{equation}\label{eq:FTOE}
\mathcal{Z}^{(n)}_{q}(O)= \mathrm{Tr}(\Pi_q(\mathrm{Tr}_{B\otimes B}(\ket{O}\bra{O}))^n)=\int_{-\pi}^{\pi}\frac{d\alpha}{2\pi}e^{-iq\alpha}Z_n(\alpha), 
\end{equation}
where $q$ is the eigenvalue of $\mathcal{Q}_A$. Thus, we have described a procedure to compute the SROE without the explicit knowledge of how the spectrum of the operator is resolved into the different charge sectors.

\subsection{SROE from the replica trick in conformal field theory (CFT)}

\subsubsection{Brief review of the replica trick for the OE in CFT}

In field theory, the reason for looking at R\'enyi entropies ---instead of focusing directly on the von Neumann entropy--- is that, for integer $n$, $\mathrm{Tr} \rho_A^{n}$ can be expressed in path-integral formalism as a partition function on an $n$-sheeted Riemann surface 
$\mathcal{R}_n$ obtained by joining cyclically the $n$ sheets along the region $A$~\cite{cc-04,cw-94}. 
One approach to compute these R\'enyi entropies is based on a particular type of twist fields in quantum field theory that are associated with the branch points of the Riemann surface $\mathcal{R}_n$.  We denote them by $\mathcal{T}_n$. Their action, in operator formalism, is defined by~\cite{cc-09,ccd-08,cd-09} 
\begin{equation}
\begin{split}
\mathcal{T}_n(x_1)\ \phi_i(x')=\phi_{i+1}(x') \mathcal{T}_n(x_1),\quad \mathcal{T}_{-n}(x_2)\ \phi_i(x')&=\phi_{i-1}(x') \mathcal{T}_{-n}(x_2),
\end{split}
\end{equation}
where $x_1$ and $x_2$ are the endpoints of the interval $A=[x_1,x_2]$ on the real axis, $x'$ is a point in $A$ and $i=1,\dots,n$ indexes the $n$ sheets modulo $n$. Here $\phi_i(z')=\mathbbm{1} \otimes \dots \otimes \phi_i(z')\dots \otimes \mathbbm{1}$ denotes any operator acting in a single copy.
By definition, the two-point function of the twist fields is the partition function on $\mathcal{R}_n$
which enters the definition of the R\'enyi entropies~\cite{cc-09}.
In conformal invariant theories the two-point function is fixed by the scaling dimension of the fields, which here is
\begin{equation}\label{eq:scaling}
  \Delta_{\mathcal{T}_{\sigma}}=\frac{c}{12}\left(n-\frac{1}{n}\right).
 \end{equation}
 We can generalise the action of the twist operators to arbitrary permutations $\sigma \in S_{n}$ of the permutation group of the $n$ copies of the field theory, namely we can define a twist operator $\mathcal{T}_{\sigma}$ as the operator with smallest possible scaling dimension such that~\cite{dubail}
\begin{equation}
\begin{split}
\mathcal{T}_{\sigma}(x_1)\ \phi_i(x')&=\phi_{\sigma(i)}(x') \mathcal{T}_{\sigma}(x_1).\\
\end{split}
\end{equation}
 If $\sigma$ has cycles of lengths $n_1+\dots + n_c=n$, then the scaling dimension of $\mathcal{T}_{\sigma}$ is 
 \begin{equation}
    \label{eq:scaling_single}
\Delta_{\mathcal{T}_{\sigma}}=\Delta_{n_1}+\dots+\Delta_{n_c},\quad \Delta_{n_i}=\frac{c}{12}\left(n_i-\frac{1}{n_i} \right),
 \end{equation}
where $c$ is the central charge of the theory. The R\'enyi entropy is given in terms of the correlator of the twist fields located at the two endpoints of $A$~\cite{cc-09,dubail}
 \begin{equation}\label{eq:state}
 S_n(\ket{\psi})=\frac{1}{1-n}\log(\bra{\psi}^{\otimes n} \mathcal{T}_{\sigma}(x_1) \mathcal{T}_{\sigma^{-1}}(x_2)\ket{\psi}^{\otimes n}),
 \end{equation}
 where $\ket{\psi}^{\otimes n}=\ket{\psi} \otimes \ket{\psi} \dots \otimes\ket{\psi} \in \mathcal{H}^n$ and $\sigma=(1,2,\dots n)$. Here $\sigma$ has a single cycle of length $n$, so it has the scaling dimension (\ref{eq:scaling_single}). The OE of an operator $O$ can be computed in a similar fashion. One has to consider $n$ replicas of $O$, $O^{\otimes n}=O \otimes O \dots \otimes O$. Then the analogous expression of Eq.~\eqref{eq:state} reads~\cite{dubail}
\begin{equation}\label{eq:sno}
S_n(O)=\frac{1}{1-n}\log \left( \frac{\mathrm{Tr}\left[ (O^{\dagger})^{\otimes n} \mathcal{T}_{\sigma}(x_1) \mathcal{T}_{\sigma^{-1}}(x_2)O^{\otimes n}\mathcal{T}_{\sigma^{-1}}(x_1) \mathcal{T}_{\sigma}(x_2)\right]}{\mathrm{Tr}((O^{\dagger})^{\otimes n}O^{\otimes n})}\right) .
\end{equation}
We now generalise this formula to include the charge operators needed for the charged moments.

\subsubsection{Method for computing the charged moments}
 
The replica trick can be adapted as follows to obtain the charged moments $Z_n(\alpha)$. The trick consists in inserting an Aharonov-Bohm flux through the multi-sheeted Riemann surface $\mathcal{R}_n$ such that the total phase accumulated by the field upon going through the entire surface is $\alpha$, similarly to what done for the standard entanglement entropy~\cite{SG17}. In terms of twist fields, this amounts to computing
\begin{equation}\label{eq:twistalpha}
Z_n(\alpha)= \frac{\mathrm{Tr}\left[ (O^{\dagger})^{\otimes n} \mathcal{T}_{\sigma}(x_1) \mathcal{T}_{\sigma^{-1}}(x_2) e^{-i\alpha Q_{A,1}} O^{\otimes n}e^{i\alpha Q_{A,1}}\mathcal{T}_{\sigma^{-1}}(x_1) \mathcal{T}_{\sigma}(x_2)\right]}{\mathrm{Tr}((O^{\dagger})^{\otimes n}O^{\otimes n})},
\end{equation}
where $Q_{A,1} = Q_1 \otimes \mathbbm{1} \otimes \dots \otimes \mathbbm{1}$ is the charge operator acting in the first replica.
Let us stress that the operator $e^{i\alpha Q_{A,1}}$ appears with opposite signs because of the definition of the charge operator in the doubled Hilbert space (see Eq.~\eqref{eq:chargeA}).

In this paper, we will apply formula (\ref{eq:twistalpha}) in the context of a spinless Luttinger liquid, which is equivalent to a $c=1$ compactified boson CFT parameterised by a coupling constant $K$ called `Luttinger parameter' see e.g. Refs.~\cite{giamarchi,tsvelik}. Crucially, in the Luttinger liquid, the $U(1)$ symmetry is generated by $Q=\frac{\sqrt{K}}{2\pi}\int_{-\infty}^\infty dx\, \partial \varphi(x)$, where $\varphi(x)$ is the boson field. Then when the charge operator is restricted to the interval $A= [x_1,x_2]$, the operators $e^{iQ_A\alpha}$ appearing in Eq.~\eqref{eq:twistalpha} can be written as 
\begin{equation}
    \label{eq:charge_end_point}
    e^{iQ_A\alpha}=e^{i\frac{\alpha \sqrt{K}}{2\pi}\int_{x_1}^{x_2} dx \partial \varphi(x)}=e^{i\frac{\alpha \sqrt{K}}{2\pi}(\varphi(x_2)-\varphi(x_1))}, 
    \end{equation}
which is a product of two vertex operators $V_\alpha(x_2) V_{-\alpha}(x_1)$ with $V_{\alpha}(x)=e^{i\frac{\alpha \sqrt{K}}{2\pi} \varphi(x)}$.
The vertex operator $V_{-\alpha}(x_1)$ can be fused with the twist field $\mathcal{T}_{\sigma^{-1}}(x_1)$ into a single `charged twist field', $\mathcal{T}_{\sigma^{-1}}^{-\alpha}(x_1) \equiv \mathcal{T}_{\sigma^{-1}}(x_1) V_{-\alpha}(x_1)$~\cite{SG17}. In other words, when one turns around the operator $\mathcal{T}_{\sigma^{-1}}^{-\alpha}(x_1)$ on the Riemann sheet, one goes from replica $j$ to replica $\sigma(j)$ and one also picks up a phase $\alpha$ when going through the first replica. The scaling dimension of the composite (or charged) twist fields reads~\cite{SG17}
\begin{equation}\label{eq:scalingK}
    \Delta_{\mathcal{T}_{\sigma}^{\alpha}}=\frac{1}{12}\left(n-\frac{1}{n} \right)+\frac{K}{n}\left(\frac{\alpha}{2\pi} \right)^2.
\end{equation}
In Section~\ref{sec:thermal} we will apply formulas (\ref{eq:twistalpha})-(\ref{eq:charge_end_point})-(\ref{eq:scalingK}) to compute the SROE of the thermal density matrix.

\subsection{Computing the SROE in free fermion chains}\label{sub:ff}
It is well established that, for the eigenstates of quadratic Hamiltonians, the entanglement entropies can be efficiently computed in terms of the eigenvalues of the correlation matrix of the
subsystem~\cite{Peschel,peschel1,peschel2}. As pointed out in Ref.~\cite{rath-22}, this strategy can be adapted to compute the charged moments of the OE, see Eq.~\eqref{eq:charged_moments}. Let us briefly review the resulting formalism here. 
The operator $O$ that we are interested in is the Gaussian density matrix $\rho$ of a free fermionic chain of length $L$. The density matrix can be diagonalised and be put in the form $\rho \propto e^{-\sum_k \lambda_kc^{\dagger}_k c_k}$, where $e^{-\lambda_k}=\frac{n_k}{1-n_k}$ with $n_k$ the occupation number of the orbital $k = 1, \dots, L$. Here $c_k^{\dagger}$ ($c_k$) creates (annihilates) a fermion in the orbital $k$; the creation/annihilation operators satisfy $\{c_k,c_{k'}^{\dagger}\}=\delta_{kk'}$.

For our purposes it is convenient to write the density matrix $\rho$ as
\begin{equation}\label{eq:rhodiag2}
\begin{split}
\rho =\bigotimes_{k=1}^{L} \frac{\ket{0}_k\bra{0}_k+e^{-\lambda_k}\ket{1}_k\bra{1}_k}{1+e^{-\lambda_k}}=\bigotimes_{k=1}^{L} [(1-n_k)\ket{0}_k\bra{0}_k+n_k\ket{1}_k\bra{1}_k],
\end{split}
\end{equation}
where $\left| 0 \right>_k$ (resp. $\left| 1 \right>_k$) are states where the orbital $k$ is empty (resp. filled),
so that by applying the vectorization trick in Eq.~\eqref{eq:vectorisation} for $\rho$, we get 
\begin{equation}
\begin{split}
    \frac{\ket{\rho}}{\sqrt{\mathrm{Tr}[\rho^2]}} =\bigotimes_{k=1}^{L} \frac{[(1-n_k)\ket{0}_k\ket{0}_{\tilde{k}}+n_k\ket{1}_k\ket{1}_{\tilde{k}}]}{\sqrt{n^2_k+(1-n_k)^2}}=\bigotimes_{k=1}^{L} \frac{[1-n_k+n_kc^{\dagger}_k\tilde{c}^{\dagger}_k] \ket{0}}{\sqrt{n^2_k+(1-n_k)^2}},
    \end{split}
\end{equation}
where the $\tilde{c}_k$ operators are copies of the $c_k$'s introduced in the vectorization process, and $\ket{0}$ is the vacuum annihilated by all the $c_k$’s and $\tilde{c}_k$'s. From this pure state, we can build the super-reduced-density-matrix $\mathrm{Tr}_{B\otimes B}(\ket{\rho}\bra{\rho})$.
The correlation matrix of the state $\ket{\rho}$ reads~\cite{rath-22,zaletel-22}
\begin{equation} \label{eq:correl_mtx}
\begin{split}
C_{kk'}=\bra{\rho} \begin{pmatrix}
c_k^{\dagger}\\
\tilde{c}_k
\end{pmatrix}
\begin{pmatrix}
c_{k'} \, \tilde{c}_{k'}^{\dagger}
\end{pmatrix}\ket{\rho}=\frac{\delta_{kk'}}{n^2_k+(1-n_k)^2}\begin{pmatrix}
& n_k^2 &n_k(1-n_k) \\
&n_k(1-n_k)  & (1-n_k)^2 \\
\end{pmatrix}.
\end{split}
\end{equation}
In the basis of $c_k, \tilde{c}_k$'s, the supercharge operator takes the form 
\begin{equation}\label{eq:charge2}
\mathcal{Q}=(\sum_k c^{\dagger}_k c_k)\otimes \mathbbm{1}-\mathbbm{1}\otimes (\sum_k\tilde{c}^{\dagger}_k \tilde{c}_k)^T. 
\end{equation}
At this point, we can compute the $2L\times 2L$ correlation matrix as 
\begin{equation}\label{eq:corrfull}
    C=\bigoplus_{k=1}^{L}C_{kk},
\end{equation}
and by doing a Fourier transform, we can write $C$ in the spatial basis.
To evaluate the charged moments in Eq.~\eqref{eq:charged_moments}, we have to restrict the supercharge operator to $\mathcal{Q}_A$ and do the Fourier transform of the correlation matrix in Eq.~\eqref{eq:corrfull} to the subspace corresponding to the subsystem $A$, of size $L_A$. Diagonalizing the latter matrix, we get $2L_A$ real eigenvalues $\xi_i$ between 0 and 1. 

Therefore, one can compute the charged moments of the reduced density matrix built from $\ket{\rho}$ in terms of the eigenvalues $\xi_i$ as~\cite{rath-22,peschel2}
\begin{equation}\label{eq:lattice}
Z_{n}(\alpha)=e^{-i\alpha (L_A)}\prod_{a=1}^{2L_A}(\xi_a^{n}e^{i\alpha}+(1-\xi_a)^{n}).
\end{equation}
Using Eq.~\eqref{eq:FTOE}, we can compute exactly the SROE for the reduced density matrix of a free fermionic chain. We will use similar techniques to compute the SROE of an operator in Heisenberg picture in Section \ref{sec:heisenberg_picture}.

%
%

\section{SROE of a thermal density matrix}
\label{sec:thermal}

In this section we focus on the thermal density matrix, $\rho_{\beta}=e^{-\beta H}$, where $\beta$ is the inverse temperature. The density matrix is vectorised, $\rho_\beta \rightarrow \ket{\rho_\beta} \in \mathcal{H}\otimes \bar{\mathcal{H}}$. For concreteness, consider for instance the free fermion chain (equivalent to the spin-$1/2$ XX spin chain via a Jordan-Wigner transformation)
\begin{equation}\label{eq:XX}
H=-\frac{1}{2}\sum_ic^{\dagger}_{i+1}c_i +\mathrm{h.c.},\quad Q=\sum_ic^{\dagger}_{i}c_i, \qquad [H,Q] = 0 .
\end{equation}
Then the charge operator $Q$ can be promoted to a `charge super-operator' $\mathcal{Q}=Q\otimes \mathbbm{1}-\mathbbm{1}\otimes Q^T$. Notice that 
$\mathcal{Q} \ket{\rho_\beta}=0$. In this section we compute the SROE of $\rho_\beta$, relying on the fact that, at low energy, the Hamiltonian~\eqref{eq:XX} corresponds to a free fermionic CFT with $c=1$, or equivalently a Luttinger liquid with Luttinger parameter $K=1$.  We first evaluate the charged moments for the $c=1$ free fermion CFT (or Luttinger liquid with $K=1$), then extend the results to the case of interacting fermions (i.e. to Luttinger liquid with $K\neq 1$), and use the results to compute the SROE. We then benchmark our analytical results against numerics.

\subsection{Charged moments for free fermion CFT}

\begin{figure}[t!]
\centering
\subfigure
{\includegraphics[width=0.88\textwidth]{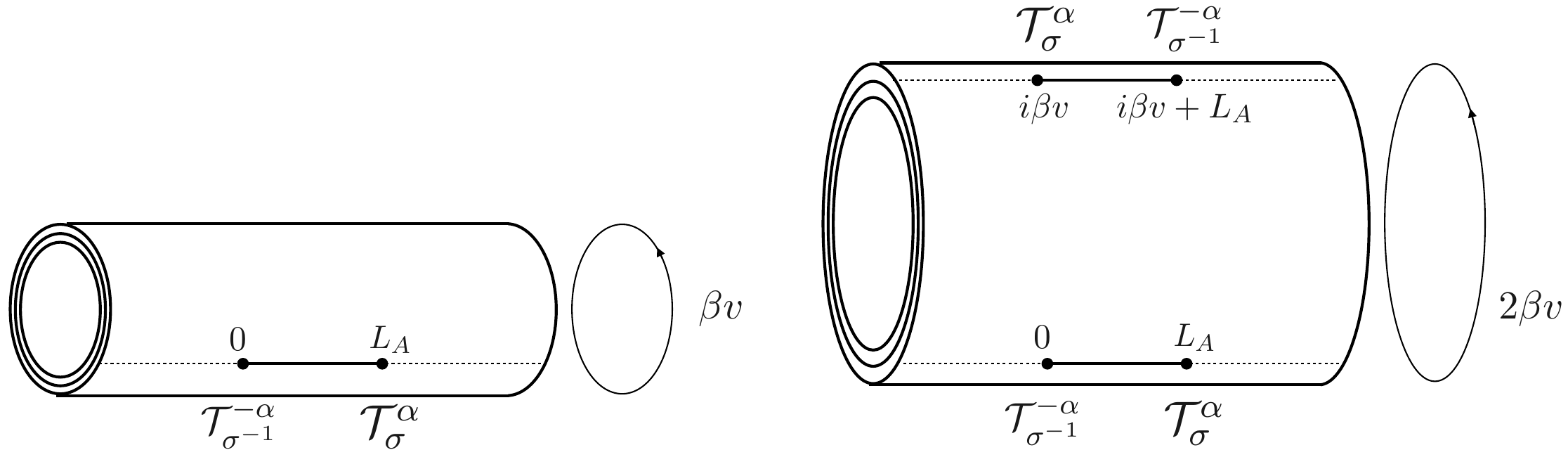}}
\caption{Left panel: The replicated surface that is used to calculate the charged moments in the setup of~\cite{SG17} on a cylinder of circumference $\beta v$. The two twist fields $\mathcal{T}^{-\alpha}_{\sigma^{-1}}$ and $\mathcal{T}^{\alpha}_{\sigma}$ are located at the endpoints of the interval $A$. Right panel: For the charged moments of the OE, there are four twist fields acting at the endpoint of the interval in $\mathcal{H}_A\otimes \bar{\mathcal{H}}_A$. }\label{fig:cylinder}
\end{figure}

As reviewed in the previous section, in CFT the charged moments can be computed as
\begin{equation}\label{eq:twist1}
Z_n(\alpha)=\frac{\mathrm{Tr}(\rho_{\beta}^{\otimes n}\mathcal{T}^{-\alpha}_{\sigma^{-1}}(x_1) \mathcal{T}^{\alpha}_{\sigma}(x_1)\rho_{\beta}^{\otimes n} \mathcal{T}^{\alpha}_{\sigma}(x_2) \mathcal{T}^{-\alpha}_{\sigma^{-1}}(x_2))}{[\mathrm{Tr}\rho_{\beta}^{\otimes 2}]^n},
\end{equation}
where we have used the charged twist fields $\mathcal{T}_\sigma^\alpha(x_1) $ that cyclically permute
the replicas of the subsystem $A$, leaving $B$ untouched as in Eq.~\eqref{eq:sno}, and are simultaneously the end points of the charge operator $Q_A = \int^{x_2}_{x_1} dx \partial\varphi(x)$ as in Eq.~(\ref{eq:charge_end_point}). 
If we view the ratio (\ref{eq:twist1}) as a correlation function of twist operators living on an infinitely long cylinder of circumference $2\beta v$, parametrised by the complex coordinate $x + iy$ with $(x, y) \in \mathbb{R} \times [0, 2\beta v]$,  then the
four charged twist operators are located at the points $0, L_A, i\beta v$ and $L_A + i\beta v$, where $L_A$ is the size of the subsystem $A$ (see Fig.~\ref{fig:cylinder}).

Therefore, the four-point function we need to calculate reads
\begin{equation}
Z_n(\alpha)=\braket{\mathcal{T}^{-\alpha}_{\sigma^{-1}}(0) \mathcal{T}^{\alpha}_{\sigma}(i\beta v) \mathcal{T}^{\alpha}_{\sigma} (L_A)\mathcal{T}^{-\alpha}_{\sigma^{-1}}(i\beta v +L_A)},
\end{equation}
where $\sigma = (1, 2, \dots, n)$ is a cyclic permutation. The scaling dimension of the composite twist field is given by Eq.~\eqref{eq:scalingK} with $K=1$.
For free fermions, this object corresponds to the evaluation of the four-point function of charged twist fields on the cylinder of
circumference $2\beta v$. The result is given in Ref.~\cite{dubail} for $\alpha = 0$, and it can be generalised straightforwardly to $\alpha \neq 0$ by replacing the scaling dimension (\ref{eq:scaling}) by (\ref{eq:scalingK}), leading to
\begin{equation}\label{eq:twist}
\log Z_n(\alpha)=-\left[ \frac{n^2-1}{3n}+\frac{\alpha^2}{\pi^2 n} \right]\log \left( \frac{2\beta v}{\pi} \tanh\left(\frac{\pi L_A}{2\beta v} \right)\right) + \log c_{n,\alpha},
\end{equation}
with $c_{n,\alpha}$ a non-universal constant which depends on the microscopic details of the model, and that we will discuss in more detail below. We implicitly assume that the lattice spacing is set equal to 1. We use this result to compute the SROE in Section~\ref{sec:SROErho}.

\subsection{Generalization to $K\neq 1$}
The result (\ref{eq:twist}) for the charged moments can be generalised to a Luttinger liquid ---or equivalently a compactified boson CFT--- with a Luttinger parameter $K \neq 1$, corresponding to interacting fermions. The scaling dimension of the composite twist field in this case changes according to Eq.~\eqref{eq:scalingK}. 

We are interested in computing $\braket{\mathcal{T}^{-\alpha}_{\sigma^{-1}}(u_1) \mathcal{T}^{\alpha}_{\sigma}(v_1) \mathcal{T}^{\alpha}_{\sigma} (u_2)\mathcal{T}^{-\alpha}_{\sigma^{-1}}(v_2)}$, with $u_1=0,v_1=L_A,u_2=i\beta v, v_2=i\beta v+L_A$. 
The calculation of that four-point function is much more complicated for $K\neq 1$ than for $K=1$, as it involves more data from the underlying CFT, including operator content and OPE coefficients (see e.g.~\cite{twist2,cct-11,twist3,twist1}). Here we exploit the result of the calculation for two intervals on the infinite line from Ref.~\cite{multicharged}, and we use the conformal transformation $w \to z = \beta v/\pi \log w$ that maps each sheet in the $w$-plane into an infinitely long cylinder of circumference $2\beta v$, to obtain the result that we need. It is convenient to introduce the cross-ratio
\begin{equation}
x=\frac{\sinh \left(\frac{\pi  (v_1-u_1)}{2 \beta  v}\right) \sinh
   \left(\frac{\pi  (u_2-v_2)}{2 \beta  v}\right)}{ \sinh\left(\frac{\pi  (v_2-u_1)}{2 \beta  v}\right)
   \sinh\left(\frac{\pi  (u_2-v_1)}{2 \beta  v}\right)}=\left[\tanh \frac{\pi L_A}{2\beta v}\right]^2 .
\end{equation}
Then the four-point function of the charged twist fields splits into a product of the four-point function of the twist fields in the plane, times the four-point function of vertex operators in the $n$-sheeted Riemann surface $\mathcal{R}_n$~\cite{multicharged},
\begin{equation}\label{eq:gen_K_new}
\begin{split}
 Z_n(\alpha) &= \braket{\mathcal{T}^{-\alpha}_{\sigma^{-1}}(u_1) \mathcal{T}^{\alpha}_{\sigma}(v_1) \mathcal{T}^{\alpha}_{\sigma} (u_2)\mathcal{T}^{-\alpha}_{\sigma^{-1}}(v_2)}\\
& = \frac{c_{n,\alpha}}{c_{n,0}} \braket{\mathcal{T}_{\sigma^{-1}}(u_1) \mathcal{T}_{\sigma}(v_1) \mathcal{T}_{\sigma} (u_2)\mathcal{T}_{\sigma^{-1}}(v_2)}  \, \braket{ V_{-\alpha} (u_1) V_\alpha (v_1) V_\alpha  (u_2) V_{-\alpha}(v_2)}_{\mathcal{R}_n} .
\end{split}
\end{equation}
Here $c_{n,\alpha}$ denote again ultra-violet non-universal constant. The factor $\braket{\mathcal{T}_{\sigma^{-1}}(u_1) \mathcal{T}_{\sigma}(v_1) \mathcal{T}_{\sigma} (u_2)\mathcal{T}_{\sigma^{-1}}(v_2)}$ is the partition function (\ref{eq:twist1}) at $\alpha=0$, while the product of vertex operators is readily evaluated~\cite{multicharged}, leading finally to
\begin{equation}\label{eq:gen_K}
\begin{split}
&\braket{\mathcal{T}^{-\alpha}_{\sigma^{-1}}(u_1) \mathcal{T}^{\alpha}_{\sigma}(v_1) \mathcal{T}^{\alpha}_{\sigma} (u_2)\mathcal{T}^{-\alpha}_{\sigma^{-1}}(v_2)}\\
&=c_{n,\alpha}\frac{Z_n(0)}{c_{n,0}}\left[\frac{4 \beta ^2 v^2}{\pi ^2} \sinh \left(\frac{\pi  (v_1-u_1)}{2 \beta  v}\right)
   \sinh \left(\frac{\pi  (v_2-u_2)}{2 \beta  v}\right)\frac{\sinh \left(\frac{\pi  (u_2-u_1)}{2 \beta  v}\right) \sinh \left(\frac{\pi 
   (v_2-v_1)}{2 \beta  v}\right)}{\sinh \left(\frac{\pi  (v_2-u_1)}{2 \beta  v}\right) \sinh \left(\frac{\pi 
   (u_2-v_1)}{2 \beta  v}\right)}\right]^{-\frac{\alpha^2 K}{2\pi^2 n}}
\\&=c_{n,\alpha}\frac{Z_n(0)}{c_{n,0}}\left[\frac{2 \beta v}{\pi}\tanh\left(\frac{L_A\pi}{2\beta v}\right)\right]^{-\frac{K\alpha^2}{\pi^2 n}}.
\end{split}
\end{equation}
The partition function $Z_n(0)$ defined in Eq.~(\ref{eq:twist1}) with $\alpha=0$ can be computed by combining the expression for the OE in Ref.~\cite{dubail} with the result for the four-point function of twist fields in Ref.~\cite{twist2}. The result is
\begin{equation}
\begin{split}
Z_n(0)=c_{n,0}\left[\frac{2 \beta v}{\pi}\tanh\left(\frac{L_A\pi}{2\beta v}\right)\right]^{-\frac{n^2-1}{3n}}\mathcal{F}_n(x),
\end{split}
\end{equation}
where $\mathcal{F}_n(x)$ is the conformal block of twist fields~\cite{twist2},
\begin{equation}\label{eq:mathcal_F}
 \mathcal{F}_n(x)=\frac{\Theta(\boldsymbol{0}|\Gamma(x)/K)\Theta (\boldsymbol{0}|\Gamma(x)K)}{\left[\Theta (\boldsymbol{0}|\Gamma(x))\right]^2}.
\end{equation}
Here $\Theta( {\bf u} | \Omega )$ is the Riemann-Siegel Theta function defined for an  $(n-1)\times(n-1)$ complex matrix $\Omega$ and an $(n-1)$-dimensional vector ${\bf u}$:
\begin{equation}
\label{eq:Siegel_Theta}
\Theta (\boldsymbol{u}|\Omega)\equiv
\sum_{\boldsymbol{m}\in\mathbb{Z}^{n-1}}
e^{i\pi\boldsymbol{m}^t
\cdot \Omega\cdot \boldsymbol{m}+2\pi i \,
\boldsymbol{m}^t\cdot \boldsymbol{u}} .
\end{equation} 
The matrix $\Gamma(x)$ in Eq.~(\ref{eq:mathcal_F}) has entries given by~\cite{twist2}
\begin{equation}\label{eq:matrix_periods}
\Gamma_{rs}(x)=\frac{2i}{n}
\sum_{l=1}^{n-1}\cos\left[\frac{2 \pi l (r-s)}{n}\right]\sin\left(\frac{\pi l}{n}\right)
\beta_{l/n}(x),\quad r, s=1, \dots,n-1,
\end{equation}
and
\begin{equation}
\label{eq:beta_def}
\beta_p(x)=\frac{I_p(1-x)}{I_p(x)},
\end{equation}
with $I_p(x)\equiv \,_2F_1(p,1-p,1,1-x)$. The conformal block $\mathcal{F}_n(x)$ is invariant under $x\mapsto 1-x$ and it is normalised such that $\mathcal{F}_n(0)=\mathcal{F}_n(1)=1$.

Our final result for the charged moment is then
\begin{equation}
    \label{eq:Znalpha_final}
    Z_n(\alpha) = c_{n,\alpha}  \left[\frac{2 \beta v}{\pi}\tanh\left(\frac{L_A\pi}{2\beta v}\right)\right]^{-\frac{n^2-1}{3n} - \frac{K \alpha^2}{\pi^2 n} }\mathcal{F}_n(x) .
\end{equation}
In the two limiting cases $L_A \gg \beta v$ and $L_A \ll \beta v$, where the four-point function factorises into a product of two two-point functions, we see that the result reduces to the expected one,
\begin{equation}\label{eq:asymp3}
    \begin{split}
    L_A \ll \beta v: &\braket{\mathcal{T}^{-\alpha}_{\sigma^{-1}}(0) \mathcal{T}^{\alpha}_{\sigma}(i\beta v) \mathcal{T}^{\alpha}_{\sigma} (L_A)\mathcal{T}^{-\alpha}_{\sigma^{-1}}(i\beta v +L_A)}\\ \simeq &\braket{\mathcal{T}^{-\alpha}_{\sigma^{-1}}(0)\mathcal{T}^{\alpha}_{\sigma} (L_A)}\braket{ \mathcal{T}^{-\alpha}_{\sigma^{-1}}(i\beta v +L_A)\mathcal{T}^{\alpha}_{\sigma}(i\beta v)}\propto
\frac{1}{L_A^{4\Delta_{\mathcal{T}_{\sigma}^{\alpha}}}}, \\
L_A \gg \beta v: &\braket{\mathcal{T}^{-\alpha}_{\sigma^{-1}}(0) \mathcal{T}^{\alpha}_{\sigma}(i\beta v) \mathcal{T}^{\alpha}_{\sigma} (L_A)\mathcal{T}^{-\alpha}_{\sigma^{-1}}(i\beta v +L_A)}\\ \simeq &\braket{\mathcal{T}^{-\alpha}_{\sigma^{-1}}(0)\mathcal{T}^{\alpha}_{\sigma} (i\beta v)}\braket{ \mathcal{T}^{-\alpha}_{\sigma^{-1}}(i\beta v +L_A)\mathcal{T}^{\alpha}_{\sigma}(L_A)}\propto
\frac{1}{\left[\frac{2\beta v}{\pi}\right]^{4\Delta_{\mathcal{T}_{\sigma}^{\alpha}}}}. 
    \end{split}
\end{equation}
These results for the charged moments have the same dependence on the parameters $\beta v, L_A$ as the one of the OE found in~\cite{dubail}, with the scaling dimension $\Delta_{\mathcal{T}_\sigma}$ replaced by $\Delta_{\mathcal{T}_\sigma^\alpha}$. Moreover, when $K=1$, $ \mathcal{F}_n(x)=1$ for all $x$ and the logarithm of Eq.~\eqref{eq:gen_K} coincides with Eq.~\eqref{eq:twist}.

\subsection{Symmetry resolution}\label{sec:SROErho}

To go from the charged moment (\ref{eq:Znalpha_final}) to the SROE, we need to take the Fourier transform with respect to $\alpha$. To do this, it is important to know the non universal constant $c_{n,\alpha}$. We start from the free fermion chain (\ref{eq:XX}), where $c_{n,\alpha}$ is known analytically, and later turn to the interacting case.

\subsubsection{SROE in the free fermion chain}
\label{sec:SROE_rho_FF}

The constant $c_{n,\alpha}$ is known in the free fermion chain (\ref{eq:XX}) from the charged moments of the reduced density matrix computed in Ref.~\cite{riccarda}. The ratio $c_{n,\alpha} / c_{n,0}$ behaves quadratically at small $\alpha$ as~\cite{riccarda}
\begin{equation}\label{eq:cnalpha}
    \frac{c_{n,\alpha}}{c_{n,0}} \, \underset{\alpha \rightarrow 0}{\simeq} \, e^{\alpha^2(2\gamma(n)-\frac{\log 2}{\pi^2 n})} ,
\end{equation}
where
\begin{equation}
\gamma(n)
= \frac{n}{4}\displaystyle \int_{-\infty}^{\infty} dw [\tanh^3(\pi n w )-\tanh(\pi n w)] \, i \,\log \dfrac{\Gamma(\frac{1}{2}+iw)}{\Gamma(\frac{1}{2}-iw)}.
\label{gamman}
\end{equation}
(This expression is real, as can be checked easily.) 
For later convenience, we define $h_n = 4\gamma(n)-\frac{2\log 2}{\pi^2 n}$ so that $c_{n,\alpha}/c_{n,0} \simeq e^{\frac{\alpha^2}{2} h_n}$. The constant $c_{n,0}$ was computed by Korepin and Jin in Ref.~\cite{korepin}:
\begin{equation}
\begin{split}\label{eq:korepin}
    c_{n,0}&=e^{2\Upsilon_n+\frac{1-n^2}{3n}\log 2} \\
    \Upsilon_n&=  n \int_{-\infty}^\infty  dw [\tanh( \pi w)-\tanh (\pi n w)]   \, i \, \log \frac{\Gamma(\frac12 +iw)}{\Gamma(\frac12 -iw)} \,.
\end{split}
\end{equation}

This leads to the following result for the charged moments in the free fermion chain:
\begin{equation}
    \label{eq:Znalpha_FF}
     Z_n(\alpha) = c_{n,0}   \left[\frac{2 \beta v}{\pi}\tanh\left(\frac{L_A\pi}{2\beta v}\right)\right]^{-\frac{n^2-1}{3n} - \frac{ \alpha^2}{\pi^2 n} } e^{ \frac{\alpha^2}{2} h_n}  .
\end{equation}
\begin{figure}[t!]
\centering
\subfigure
{\includegraphics[width=0.495\textwidth]{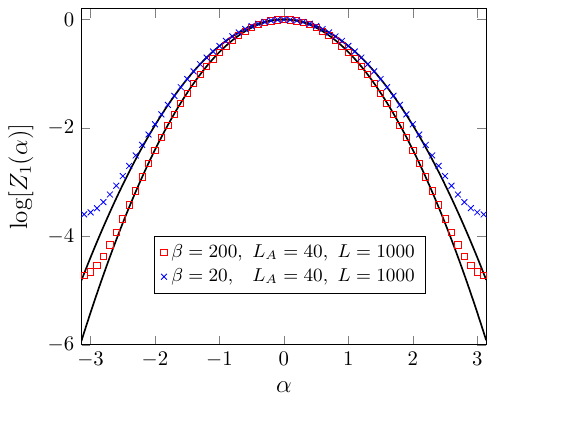}}
\subfigure
{\includegraphics[width=0.495\textwidth]{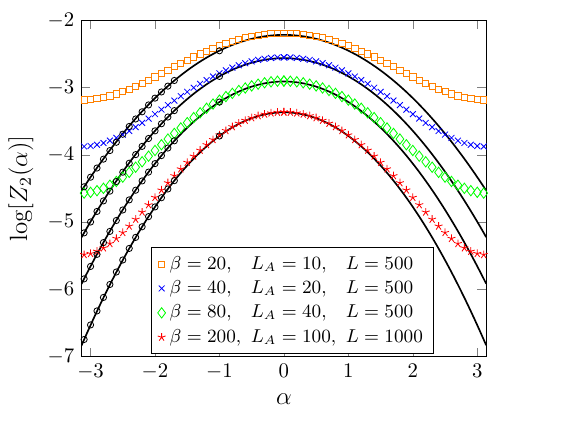}}
\caption{Numerical check of formula~\eqref{eq:Znalpha_FF} for free fermions on the lattice with dispersion $\varepsilon(k) = - \cos k$, for two different values of $n$. The Fermi velocity is $v=1$. We take a chain of $L$ sites at inverse temperature $\beta$ and cut an interval of length $L_A$. The black symbols represent the extrapolated data for fixed $L_A/\beta=0.5$.}\label{fig:extr}
\end{figure}
We check the result \eqref{eq:Znalpha_FF} against numerics in Fig.~\ref{fig:extr}. The symbols represent the numerical data obtained through exact lattice computations, using the expression for the charged moments described in Section~\ref{sub:ff}. The data present finite size corrections, which become larger as $n$ and $\alpha$ increase. In the right panel, in order to achieve the correct scaling limit, we have computed $\log Z_n(\alpha)$ at fixed $\alpha, n=2$ and for fixed $L_A/\beta=0.5$. The
leading corrections to the scaling behave as $L_A^{-2(1-\alpha/\pi)/n}$, $L_A^{-2(1+\alpha/\pi)/n}$, $L_A^{-2/n}$, $L_A^{-2/n(2-\alpha/\pi)}$ (see also~\cite{bc-21} for a similar analysis). We 
then perform a fit of the finite $L_A$ data (fox fixed $\alpha$, $n=2$), keeping the first two power-law corrections and extrapolating at $L_A \to \infty$. The data obtained following this procedure are reported as black symbols in the right panel of Fig.~\ref{fig:extr}.

\begin{figure}[ht]
\centering
\subfigure
{\includegraphics[width=0.495\textwidth]{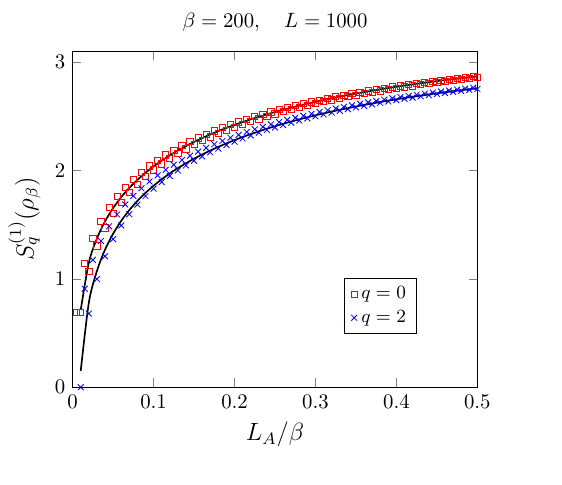}}
\subfigure
{\includegraphics[width=0.495\textwidth]{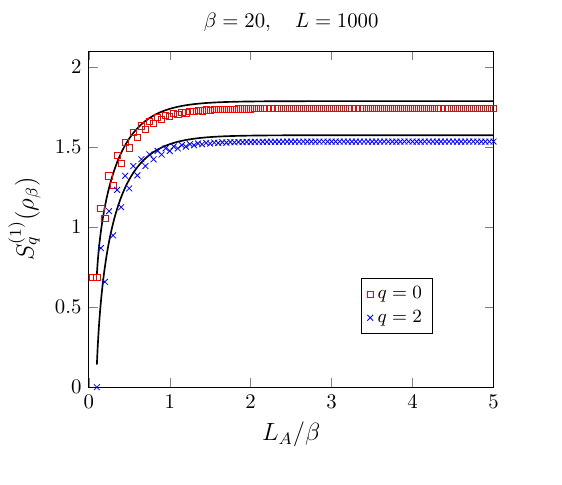}}
\subfigure
{\includegraphics[width=0.495\textwidth]{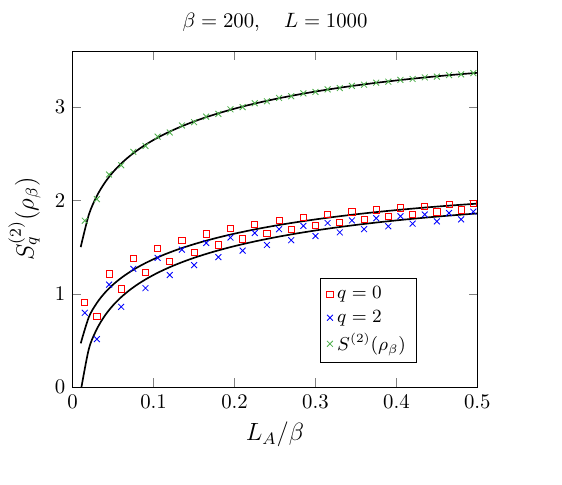}}
\subfigure
{\includegraphics[width=0.495\textwidth]{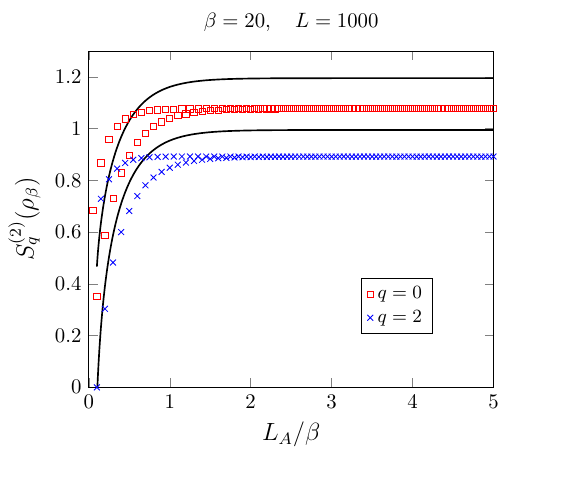}}
\caption{Eq.\eqref{eq:lattice} (symbols) vs Eq.~\eqref{eq:SP-SRRE-v2Order} (solid lines) for $\beta=200, 20$, $n=1$ (upper panel), $n=2$ (lower panel). For $n=2$, we also check the result for the total OE (green line). We observe that as the temperature increases, the finite size corrections become more important. }\label{fig:2}
\end{figure}

Plugging the expression~\eqref{eq:Znalpha_FF} of the charged moments in the Fourier transform~\eqref{eq:FTOE} and applying the saddle-point approximation for $\beta, L_A \gg 1$, we get 
\begin{multline}
\label{eq:SP-SRRE-v2Order}
S^{(n)}_{q}(\rho_{\beta})=
S^{(n)}(\rho_{\beta})-\frac{1}{2} \log \left(\frac{4}{\pi} \log \delta_n \left( \frac{2\beta v}{\pi } \tanh\left(\frac{\pi L_A}{2\beta v} \right)\right) \right)+ \frac{\log n}{2(1-n)}\\-\frac{\pi^4n(h_1-nh_n)^2}{4(1-n)^2\left(\log \left( \frac{2\beta v}{\pi } \tanh\left(\frac{\pi L_A}{2\beta v} \right)\right)\right)^2}+ 
\\ +
q^2n \pi^4\frac{h_1-nh_n}{4(1-n)\left(\log  \left( \frac{2\beta v}{\pi } \tanh\left(\frac{\pi L_A}{2\beta v} \right)\right)  \kappa_n\right)^2} + o(\log^{-2} \left( \beta \tanh\left(\frac{\pi L_A}{2\beta v} \right)\right)),
\end{multline}
where 
\begin{equation}
\label{eq:deltan}
\log \delta_n=-\dfrac{\pi^2 n  (h_n-h_1)}{(1-n)},\qquad \log \kappa_n=-\pi^2\frac{(h_1+n h_n)}{2},
\end{equation}
and $S^{(n)}(\rho_\beta)$ is the total OE 
\begin{equation}
S^{(n)}(\rho_{\beta}) = \frac{n+1}{3n} \log \left( \frac{4\beta v}{\pi } \tanh\left(\frac{\pi L_A}{2\beta v} \right)\right) +\frac{2}{1-n}\Upsilon_n .
\end{equation}
This formula is also valid for the symmetry resolved von Neumann OE taking properly the limits of the various pieces when $n\to 1$.  The expression~\eqref{eq:SP-SRRE-v2Order} is checked against numerics in the free fermion chain in Fig.~\ref{fig:2}. The small deviations between data and analytics are a consequence of the finite size corrections already present in the charged moments (Fig.~\ref{fig:extr}).

\subsubsection{Interacting case}

For $K \neq 1$, we do not know a microscopic model where the non-universal constant $c_{n,\alpha}$ is known. Nevertheless, we can generalise the previous results, assuming that $c_{n,\alpha}$ behaves smoothly as a function of $\alpha$ near $\alpha = 0$, i.e.
\begin{equation}
    c_{n,\alpha} = c_{n,0}  - \frac{\alpha^2}{2} c_{n,0} h_n + o(\alpha^2) = c_{n,0} e^{ \frac{\alpha^2}{2} h_n }  + o(\alpha^2)
\end{equation}
for some constants $c_{n,0}$ and $h_n$ that depend on the microscopic model and are, in general, not equal to the ones in the free fermion chain. By repeating the same steps as before, the SROE reads
\begin{multline}
\label{eq:SP-SRRE-v2Order_K}
S^{(n)}_{q}(\rho_{\beta})=
S^{(n)}(\rho_{\beta})-\frac{1}{2} \log \left(\frac{4K}{\pi} \log \delta_n \left( \frac{2\beta v}{\pi } \tanh\left(\frac{\pi L_A}{2\beta v} \right)\right) \right)+ \frac{\log n}{2(1-n)}\\-\frac{\pi^4n(h_1-nh_n)^2}{4K^2(1-n)^2\left(\log \left( \frac{2\beta v}{\pi } \tanh\left(\frac{\pi L_A}{2\beta v} \right)\right)\right)^2}+ 
\\ +
q^2n \pi^4\frac{h_1-nh_n}{4K^2(1-n)\left(\log  \left( \frac{2\beta v}{\pi } \tanh\left(\frac{\pi L_A}{2\beta v} \right)\right)  \kappa_n\right)^2} + o\Big(\log^{-2} \left( \beta \tanh\left(\frac{\pi L_A}{2\beta v} \right)\right)\Big).
\end{multline}

\subsubsection{Discussion}

To get insight into the meaning of Eqs.~\eqref{eq:SP-SRRE-v2Order} and \eqref{eq:SP-SRRE-v2Order_K}, we study the following two asymptotic regimes, which are valid for any value of $K$:
\begin{equation}\label{eq:asymp4}
    \begin{split}
    L_A \ll \beta v:& \, S^{(n)}_{q}(\rho_{\beta})=\frac{n+1}{3n} \log L_A
-\frac{1}{2}\log(K \log L_A)+O(1),\\
L_A \gg \beta v:& \, S^{(n)}_{q}(\rho_{\beta})= \frac{n+1}{3n}\log\frac{2\beta v}{\pi}-\frac{1}{2}\log(K\log\frac{2\beta v}{\pi})+O(1), 
    \end{split}
\end{equation}
We find that the leading order term corresponds to the total OE, which diverges logarithmically in $L_A$
at low temperatures, i.e. when the reduced density matrix $\rho_A$ is very close to the one of the ground state, while it is bounded (in $L_A$) at finite temperature $\beta$. This is the main striking difference with respect to the usual entanglement, which at finite temperature has an extensive behaviour. The practical consequence of this result is that $\rho_{\beta}$ can be efficiently represented by an MPO~\cite{dubail}.
Then $S^{(n)}_{q}(\rho_{\beta})$ presents a double logarithmic correction in $L_A$ at low temperature, while it remains bounded at finite temperature, with a double logarithmic correction depending on the inverse
temperature $\beta$. Finally, the fact that the leading order term of the SROE coincides with the total OE resembles the entanglement equipartition for the symmetry resolution of a $U(1)-$invariant theory~\cite{sierra}, and it can be traced back to the conformal invariance of the system also in this case.

\section{SROE of local operators in Heisenberg picture in free fermion chains}
\label{sec:heisenberg_picture}

The problem of computing the OE of local operators evolving in the Heisenberg picture, $\phi(t)=e^{iHt}\phi e^{-iHt}$ ---also called `local operator entanglement' by Kos, Bertini and Prosen~\cite{bruno1}---, has attracted some attention recently~\cite{fleischhauer,dubail,jonaynahum,alba1,alba2,bruno1,bruno2, pavel,tagliacozzo}. In this section we study the symmetry resolution of the OE of a local operator in Heisenberg picture in the XX chain with periodic boundary conditions, which is mapped to the free fermion chain (\ref{eq:XX}) by the Jordan-Wigner transformation. Following Refs.~\cite{Prosen2007,Pizorn2009,dubail}, we  distinguish initial operators that are local in terms of the fermions (such as $c_x^\dagger$ or $c^\dagger_{x} c_x$) from operators that are attached to a Jordan-Wigner string.

\subsection{SROE of creation/annihilation operators and other local operators}
The goal of this section is to study the SROE of $c_x^\dagger(t),\sigma^z_{x}(t)$. In the former case, we take advantage of the form in which Eq.~\eqref{eq:OABgeneral} can be written, while in the latter we exploit the connection between the charged moments of the super-density-matrix and the correlation matrix, similarly to what has been done in Section~\ref{sub:ff}.
\subsubsection{SROE of creation operator $c_x^\dagger(t)$}
We start by studying the SROE of a single creation operator $c_x^\dagger(t)$ (the analysis for $c_x(t)$ can be performed in a similar way). We observe that $[Q,c_x^\dagger(t)]=c_x^\dagger(t)$, with $Q=\sum_j c_j^\dagger c_j$, so $q_O=1$ in Eq.~\eqref{eq:OABgeneral}. We choose $x=0$ and we can use the Fourier transform, for a ring of $L$ sites,
\begin{equation}\label{eq:F_c}
c^{\dagger}_0(t)=\sum_jU^*_{j0}(t)c^{\dagger}_m,\quad U_{0j}(t)=\frac{1}{L}\sum_ke^{ikj}e^{it\cos(k)}.
\end{equation}
In the thermodynamic limit, the matrix elements $U_{0m}(t)$ of the unitary evolution
operator can be written in terms of Bessel functions, $J_m(t)$. We can directly recast $c^{\dagger}_0(t)$ into the form of Eq. \eqref{eq:OABgeneral}, which reads
\begin{equation}
   c^{\dagger}_0(t)=\left(\sum_{m\leq 0} J_m(t)\right)\otimes\mathbbm{1}_B +\mathbbm{1}_A\otimes \left(\sum_{m> 0} J_m(t)\right).
\end{equation}
This implies that there are only two possible charge sectors, $q=1,0$ and we can also directly write down the Schmidt values $\lambda_m(0)=\lambda_m(1)=J_m(t)$.
At large times, the Bessel functions satisfy
\begin{equation}\label{eq:Bessel}
\sum_{m\leq 0} J_m^2(t) \underset{t \rightarrow \infty}{\longrightarrow} \frac{1}{2} .
\end{equation}
Combining Eq.~\eqref{eq:pqsroe} and~\eqref{eq:Bessel}, we get the following result at large time:
\begin{equation}
    p(0)=p(1)=1/2,\qquad S_0^{(n)}(c^{\dagger}_0(t))=S_1^{(n)}(c^{\dagger}_0(t))=0
\end{equation}
and 
\begin{equation}
\displaystyle \sum_q p(q) S_{q}(c^{\dagger}_0(t)- \displaystyle \sum_q p(q) \log p(q)=2\frac{\log 2}{2}=\log 2.
\end{equation}
This implies that the SROE vanishes in all the charge sectors and the total OE is given only by the number/fluctuation entanglement (the second term of Eq.~\eqref{eq:SvN}) i.e. by the probability of finding $q=1,0$ as an outcome of $\mathcal{Q}_A$.

\subsubsection{SROE of local density operator $c_x^\dagger (t) c_x(t)$}

We now consider a less trivial example, which is the symmetry resolution of a pair of creation and annihilation operators. In the spin language (up to additive constants) this corresponds to
\begin{equation}
c^{\dagger}_x c_x \leftrightarrow \sigma^z_x.
\end{equation}
Let us focus on $\sigma^z_0(t)$. From the state-operator correspondence, we want to study the entanglement due to the state $c^{\dagger}_0(t) \tilde{c}^{\dagger}_0(t)\ket{0}$. 
Before reporting the details of the computations, we report the main result here. We observe that there are 3 non trivial charge sectors, $q=0,\pm 1$, where
\begin{equation}\label{eq:oseesz}
p(0)=2 p(\pm 1)=\frac{1}{2}\quad S^{(n)}_{q=0}(\sigma^z_0(t))=\log 2, \quad S_n(q\neq 0)=0,
\end{equation}
and 
\begin{equation}
\displaystyle \sum_q p(q) S_{q}(\sigma^z_0(t))- \displaystyle \sum_q p(q) \log p(q)=\frac{\log 2}{2}+\frac{\log 2}{2}+2\frac{\log 4}{4}=2\log 2,
\end{equation}
i.e. Eq.~\eqref{eq:SvN} is satisfied. Therefore, differently from what we found before for one single creation operator $c^{\dagger}_0(t)$, there exist one charge sector $q=0$ where the SROE is different from 0 and both terms in Eq.~\eqref{eq:SvN} contribute to the total OE.

Given the simplicity of the result, it might be possible to find it following the logic of the previous subsection about $c^{\dagger}(t)$. However, as an illustration of the techniques used in Section \ref{sub:ff}, we follow an alternative path. First of all, we need to compute 
\begin{equation}\label{eq:cdaggerc}
\bra{0}\tilde{c}_0(t)c_0(t)c^{\dagger}_nc_mc^{\dagger}_0 (t)\tilde{c}^{\dagger}_0(t)\ket{0}.
\end{equation}
Therefore, plugging Eq.~\eqref{eq:F_c} into Eq.~\eqref{eq:cdaggerc}, we get
\begin{equation}
\bra{0}\tilde{c}_0(t)c_0(t)c^{\dagger}_nc_mc^{\dagger}_0 (t)\tilde{c}^{\dagger}_0(t)\ket{0}=U_{0n}(t)U^*_{m0}(t).
\end{equation}
By repeating the same computations for $c^{\dagger}_n\tilde{c}^{\dagger}_m$, $\tilde{c}_nc_m$, $\tilde{c}_n\tilde{c}^{\dagger}_m$, we can write down the $2L_A \times 2L_A$ correlation matrix for a subsystem $(-L_A+1,0)$ as
\begin{equation}\label{eq:corr1}
C_A(t)=\begin{pmatrix}
&\mathcal{C}(t) & 0_{L_A}\\
 & 0_{L_A} & \mathbbm{1}_{L_A}-\mathcal{C}(t) \\
\end{pmatrix},
\end{equation}
where $\mathcal{C}_{nm}(t)=i^{m-n}J_m(t)J_n(t)$, while $0_{L_A}$ and $ \mathbbm{1}_{L_A}$ denote the $L_A\times L_A$ null and identity matrix, respectively. 
 In the scaling limit $t, L_A \to \infty$, we can compute $\mathrm{Tr}(2C_A(t)-1)^j$, observing that 
\begin{equation}
\mathrm{Tr}(2C_A(t)-1)^{2j}=2L_A-2, \quad \mathrm{Tr}(2C_A(t)-1)^0=2L_A,\quad  \mathrm{Tr}(2C_A(t)-1)^{2j-1}=0.
\end{equation} 
This can be understood as follows: first of all, because of the construction of $C_A(t)$ with the blocks $\mathcal{C}(t)$ and $1-\mathcal{C}(t)$, $\mathrm{Tr}(2C_A(t)-1)^j$ vanishes for any odd values of $j$. For the even powers, we can start by studying the case $j=2$. We restrict to the upper $L_A \times L_A$ block and we compute the following trace (the trace of the square of the lower block is the same)
\begin{equation}\label{eq:tracej2}
\mathrm{Tr}(2\mathcal{C}(t)-1)^2=\sum_{kj}(2J_k(t)J_j(t)i^{k-j}-\delta_{kj})(2J_k(t)J_j(t)i^{j-k}-\delta_{kj}).
\end{equation} 
Therefore, Eq.~\eqref{eq:tracej2} and \eqref{eq:Bessel} give
\begin{equation}
4\sum_{kj}J^2_k(t)J^2_j(t)-4\sum_jJ^2_j(t)+\sum_j 1=L_A-1.
\end{equation}
The generalisation to higher powers of $j$ follows the same logic, since all the powers involving terms like 
$
\sum_j J_j^{2k}(t) $ vanish for $k>1$ in the scaling limit.

\begin{figure}[t!]
\centering
\subfigure
{\includegraphics[width=0.48\textwidth]{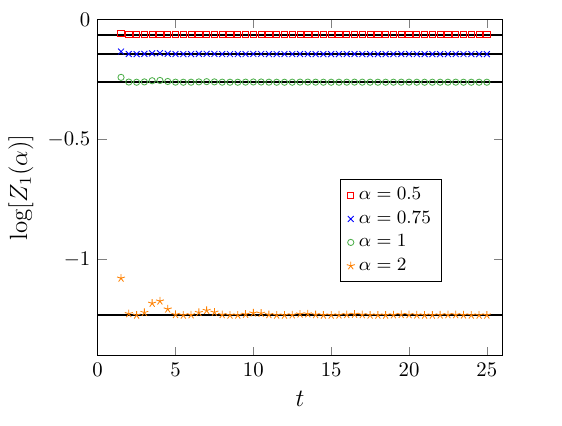}}
\subfigure
{\includegraphics[width=0.48\textwidth]{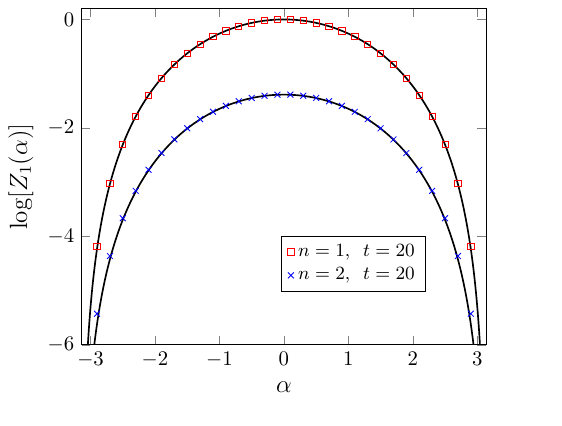}}
\caption{Logarithm of the charged moments for the operator $\sigma_0^z(t)$ in the tight-binding model (see Eq.~\eqref{eq:XX}) for a bipartition $[-L_A,0]\cup [1,L_A]$, $L_A=30$, with periodic boundary conditions.
The plots are at fixed $\alpha$ as function of time (left) and at fixed time as function of $\alpha$ (right). 
The symbols are the numerical data coming from the evaluation of the eigenvalues of the correlation matrix in Eq.~\eqref{eq:corr1} while the solid line represents Eq.~\eqref{eq:sigmaz}.}\label{fig2:sigmaz}
\end{figure}

We can now come back to our main focus of the section, i.e the evaluation of the charged moments built from $\sigma^z_0(t)$. The charged moments in Eq.~\eqref{eq:lattice} can be rewritten using~\cite{pbc-21,fc-08}
\begin{equation}\label{eq:lattice2}
\begin{split}
h_{n,\alpha}(x)=&\log \left[ \left( \frac{1+x}{2} \right)^n e^{i\alpha}+\left(\frac{1-x}{2} \right)^n \right]=\sum_{m=0}^{\infty}s_{n,\alpha}(m)x^m,\\
\log Z_n(\alpha)=& -i\alpha L_A+\sum_{m=0}^{\infty}s_{n,\alpha}(m)\mathrm{Tr}(2C_A(t)-1)^m.
\end{split}
\end{equation}
The coefficients $s_{n,\alpha}(m)$ correspond to the function $h_{n,\alpha}(x)$ evaluated in
certain simple points:
\begin{equation}
\begin{split}
s_n(0)=&h_{n,\alpha}(0)=(1-n)\log 2+\log \cos \frac{\alpha}{2}+i\frac{\alpha}{2},\\
\sum_{m=1}^{\infty}s_{n,\alpha}(2m)=&\left(\frac{h_{n,\alpha}(1)+h_{n,\alpha}(-1)}{2}-s_{n,\alpha}(0) \right)=(n-1)\log 2-\log \cos \frac{\alpha}{2}.
\end{split}
\end{equation}
It follows that 
\begin{equation}\label{eq:sigmaz}
\log Z_n(\alpha)= -i\alpha L_A+s_n(0)(2L_A)+2(L_A-1)\sum_{m=1}^{\infty}s_{n,\alpha}(2m)=2(1-n)\log 2+2\log \cos \frac{\alpha}{2}.
\end{equation}
In Fig.~\ref{fig2:sigmaz} we compare the prediction~\eqref{eq:sigmaz} for the logarithm of the charged moments $\log Z_n(\alpha)$ with the exact lattice computations done using the correlation matrix in Eq.~\eqref{eq:corr1}. The dependence of $Z_n(\alpha)$ in $t$ and $\alpha$ is perfectly reproduced by the exact result.

\begin{figure}[t!]
\centering
\subfigure
{\includegraphics[width=0.48\textwidth]{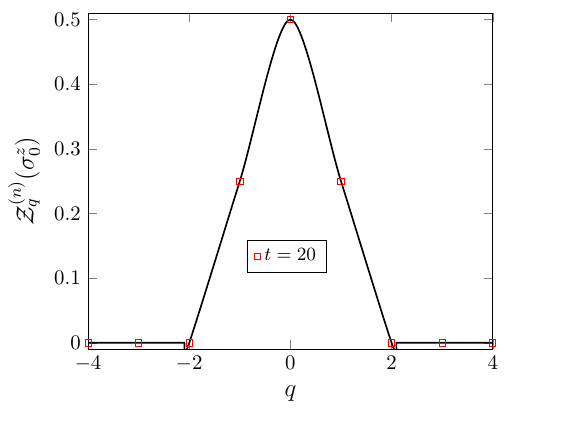}}
\subfigure
{\includegraphics[width=0.48\textwidth]{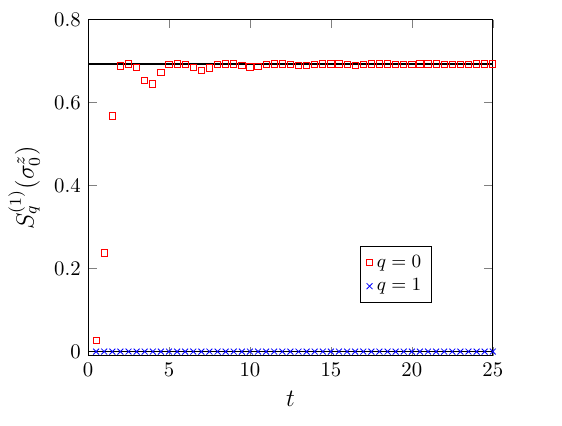}}
\caption{
Symmetry resolution of the OE for the operator $\sigma_0^z(t)$.
Left panel: probability of finding $q$ as an outcome of the measurement of $\mathcal{Q}_A$. Right panel: SROE as function of $t$ at fixed $q$. 
From Eq.~\eqref{eq:oseesz}  only the $q=0$ charge sector has a non-vanishing entropy which saturates to the constant value $\log2$ .}\label{fig2:sigmaz2}
\end{figure}

By doing the Fourier transform in Eq.~\eqref{eq:FTOE}, we can compute
\begin{equation}
\mathcal{Z}^{(n)}_{q}(\sigma^z_0(t))=2^{2(1-n)}\int_{-\pi}^{\pi}\frac{d\alpha}{2\pi}e^{-iq\alpha}Z_n(\alpha)=2^{2(1-n)}\frac{\sin(\pi q)}{2\pi(q-q^3)}.
\end{equation}
We recognise that $\mathcal{Z}^{(n)}_{q=0}(\sigma^z_0(t))=\frac{2^{2(1-n)}}{2}$, $\mathcal{Z}^{(n)}_{q=\pm 1}(\sigma^z_0(t))=\frac{2^{2(1-n)}}{4}$, otherwise $\mathcal{Z}^{(n)}_{q}(\sigma^z_0(t))=0$. This has been checked in the left panel of Fig.~\ref{fig2:sigmaz2}. Finally, we deduce that the SROE behaves as reported in Eq.~\eqref{eq:oseesz}: The fact that the OE is different from $0$ only in one charge sector is a strong violation of the equipartition that we have found in Section~\ref{sec:thermal}.
This lack of equipartition is a direct consequence of the fact that the operator entanglement stays always finite. 

\subsection{OE of a Jordan-Wigner string}
We now turn to the calculation of the SROE of a Jordan-Wigner string in the tight-binding chain with periodic boundary conditions, i.e.
\begin{equation}
JW_x(t)\equiv \prod_{y\leq x}(1-2c^{\dagger}_y(t)c_y(t)).
\end{equation} 
We place the endpoint of the string at the origin without loss of generality.
Therefore, we are interested in the correlation matrix built from 
\begin{equation}
JW_0(t)\to \ket{JW_0(t)}=\prod_{y \leq 0}(1-c^{\dagger}_y(t)\tilde{c}^{\dagger}_y(t))\ket{0},
\end{equation}
by replacing $c(t)$ with $\tilde{c}^{\dagger}(t)$ which anticommutes with all the $c$’s. It is convenient to extract the time dependence as 
\begin{equation}
\ket{JW_0(t)}=e^{it(H\otimes \mathbbm{1}-\mathbbm{1}\otimes H)}\ket{JW_0(0)},
\end{equation}
and to perform a Bogoliubov transformation (exactly as done for the total OE in~\cite{dubail}),
\begin{equation}
\begin{pmatrix}
b^{\dagger}_x\\
d_x
\end{pmatrix}=\frac{1}{\sqrt{5}}\begin{pmatrix}
 1 & -1 \\
1 &1 \\
\end{pmatrix}
\begin{pmatrix}
c^{\dagger}_x \\
\tilde{c}^{\dagger}_x
\end{pmatrix}.
\end{equation}
Thus, the initial state becomes
\begin{equation}\label{eq:DW_conn}
\ket{JW_0(t)}=\frac{1}{5}\prod_{y\leq 0}d_y^{\dagger}b_y^{\dagger}\ket{0_{d,b}},
\end{equation}
where now $\ket{0_{d,b}}$ is the vacuum for the $d$ and $b$ modes. This transformation does not do anything to the Hamiltonian $H\otimes \mathbbm{1}-\mathbbm{1}\otimes \tilde{H}$ and to the charge operator $Q\otimes  \mathbbm{1}-\mathbbm{1}\otimes Q^T$ in Eq.~\eqref{eq:charge2}, which already were diagonal. However, due to this simplification on the initial state, we can compute the correlation matrix as
\begin{equation}
C_A(t)=\begin{pmatrix}
&\mathcal{C}(t) & 0_{L_A}\\
 & 0_{L_A} & \mathbbm{1}_{L_A}-\mathcal{C}(t), \\
\end{pmatrix}
\end{equation}
where $\mathcal{C}_{nm}(t)=\sum_{y=-L_A}^0i^{m-n}J_{m-y}(t)J_{n-y}(t)$~\cite{alba-14}. 
Rewriting the initial state as in Eq.~\eqref{eq:DW_conn} also allows us to have an analytical insight on the time evolution of the SROE, following the observations done in~\cite{dubail}. Indeed, Eq.~\eqref{eq:DW_conn} is a product state $\ket{\mathrm{DWIS}}_b\otimes \ket{\mathrm{DWIS}}_d=\prod_{y\leq 0}d_y^{\dagger}\ket{0_{d}}\otimes \prod_{y\leq 0}b_y^{\dagger}\ket{0_{b}}$. The two independent
initial states for the $b$’s and the $d$’s operators are domain-wall initial states (DWIS). For a bipartition $A \cup B=(-L_A,0) \cup (0,L_A)$ and in the scaling limit $L_A \to \infty$, the charged moments of the JW string are exactly twice the charged moments of the domain wall computed in~\cite{stefano}, i.e.
\begin{equation}\label{eq:DWcm}
\log Z_n(\alpha)=-\left[ \frac{n^2-1}{3n}+\frac{\alpha^2}{\pi^2 n} \right]\log 4t+2\alpha^2\gamma(n).
\end{equation}

\begin{figure}[t!]
\centering
\subfigure
{\includegraphics[width=0.485\textwidth]{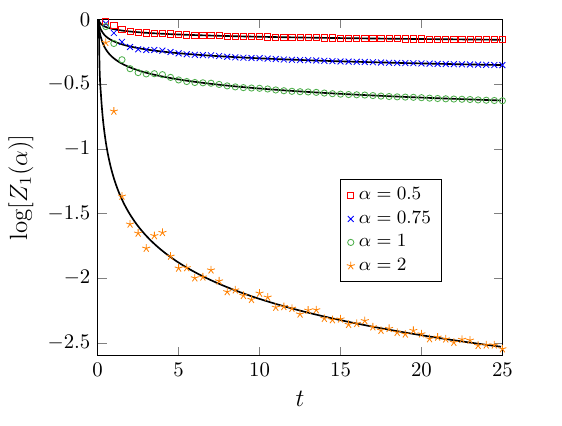}}
\subfigure
{\includegraphics[width=0.48\textwidth]{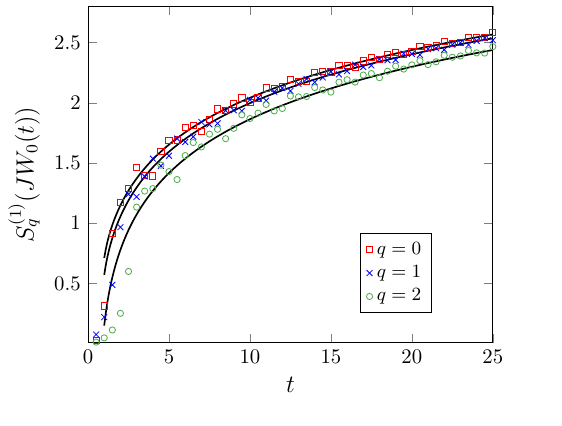}}
\caption{Left panel: logarithm of the charged moments for the Jordan-Wigner string $JW_0(t)$  in the tight-binding model for a bipartition $[-L_A+1,0]\cup [1,L_A]$, $L_A=30$, with periodic boundary conditions. The solid line is Eq.~\eqref{eq:DWcm}, which also takes into account the exact knowledge of the non-universal constants for this system. The SROE of this operator is studied in the right panel: the comparison between numerics and  the analytical prediction in Eq.~\eqref{eq:DW} is quite good.}\label{fig:sigmax}
\end{figure}

Let us remark that the knowledge of the non-universal constants from~\cite{riccarda} allows us to benchmark our analytical prediction against the exact lattice computations without fitting any parameter, as we have done in the left panel of Fig.~\ref{fig:sigmax}. By doing a Fourier transform and applying the saddle point approximation in the limit $t \to \infty$, we get
\begin{multline}
\label{eq:DW}
S_{q}^{(n)}(JW_0(t))=
S^{(n)}(JW_0(t))-\frac{1}{2} \log \left(\frac{4}{\pi} \log (2t \delta_n)\right)+ \frac{\log n}{2(1-n)}-\frac{\pi^4n(h_1-nh_n)^2}{4(1-n)^2(\log (2t))^2}+ 
\\ +
q^2n \pi^4\frac{h_1-nh_n}{4(1-n)\left(\log  (2t\kappa_n)\right)^2} + o(\log^{-2} t),
\end{multline}
where the constants $h_n$, $\delta_n$, $\kappa_n$ have been defined in Eq.~\eqref{eq:deltan} and after Eq.~\eqref{eq:cnalpha}, and 
\begin{equation}
S^{(n)}(JW_0(t))=\frac{n+1}{3n}\log (4t)+\frac{2}{1-n}\Upsilon_n.
\end{equation}
This is just the double of the entanglement entropy following a quench from a domain wall state \cite{stefano}. 
Such a  result is not surprising since it was already observed in~\cite{dubail} that the total OE of the JW string is exactly twice the entanglement entropy of a domain wall~\cite{dsvc-16} and we find that the same is true in each symmetry sector. 
Hence, the equipartition of OE in the symmetry sectors is asymptotically restored with an asymptotic  correction of the form $q^2/\log^2(t)$, with a non-trivial prefactor depending on non-universal quantities. 

The results of this and of the former subsection suggest that the breaking of the equipartition is a feature directly related to the finiteness of operator entanglement. The latter takes place for operators with a finite support (like $\sigma_z$, i.e.  $c^{\dagger}_xc_x$) while extended operators, like the Jordan-Wigner strings, have diverging (in $t$) operator entanglement and restore equipartition for large times.

%
%

\section{Conclusions}
\label{sec:conclusion}

In the light of the examples studied so far, we can now draw our general conclusions on symmetry resolved operator entanglement. We have used the definition of the SROE introduced in~\cite{rath-22}, which quantifies the OE of a $U(1)$ symmetric operator in a given charge sector and we have analysed in detail three operators. The thermal density has been studied with the twist field formalism and we found that it satisfies the operator area law, i.e. it has a bounded SROE in every charge sector and it displays equipartition, meaning that at leading order in the subsystem size it does not depend on the charge. 
For free fermion Hamiltonians, to evaluate the SROE of a local operator (in terms of the creation and annihilation operators) evolving in Heisenberg picture, we have exploited the knowledge of the SROE in terms of the correlation functions.
We found that the local density operator
obeys the area law at any time step and strongly violates the equipartition, while the SROE  a Jordan-Wigner string grows logarithmically in time and obeys the equipartition of the entanglement. 
This remarkable difference between the three operators might be traced back to the fact that the SROE of local operators remains always finite, contrarily to the Jordan-Wigner string 
 (that diverges for large $t$) or the thermal density matrix (that diverges for large $\beta$). The rationale appears to be that operators with
a local spatial support have an operator entanglement that does not diverge and cannot satisfy equipartition.  

We can now think of possible future directions that one could investigate about the OE and its eventual symmetry resolution. In a previous work about the same subject~\cite{rath-22}, the authors have studied the SROE of a reduced density matrix after a quantum quench~\cite{dubail,Wang2019barrier,Reid2021Entanglement}, and it would be interesting to study how the entanglement barrier is affected by a non-unitary time evolution, due, for example, to the effect of local measurements in the dynamics. Another possible direction is studying the connection between the OE and the reflected entropy~\cite{reflected,reflected2}, by applying the techniques described in this manuscript to analyse the symmetry resolution also in this context (see also \cite{gillesOE} for the first steps towards this direction). This example would also gain insights about the holographic dual of the SROE when $O=\sqrt{\rho}$.
Similarly one would like to explore the connection of our results with symmetry resolution of the computable cross norm negativity \cite{yl-23,gillesOE} which is another measure of entanglement in mixed states. 
A last interesting point concerns the interplay of OE and symmetries when the latter are broken; this problem can be studied generalising to the operatorial level the recently introduced entanglement asymmetry \cite{amc-22,amvc-23,fac-23,bkcc-23}.  

\section*{Acknowledgements}
We thank Filiberto Ares, Giuseppe Di Giulio, Aniket Rath and Vittorio Vitale for useful discussions.
PC acknowledges support from ERC under Consolidator grant number 771536 (NEMO). JD acknowledges support from the Agence Nationale de la Recherche through ANR-20-CE30-0017-
01 project ‘QUADY’ and ANR-22-CE30-0004-01 project
‘UNIOPEN’. SM thanks the Caltech Institute for Quantum Information and Matter and the Walter Burke Institute for Theoretical Physics at Caltech.
All authors acknowledge the 
hospitality of the Simons Center for Geometry and Physics in Stony Brook during the program `Fluctuations, Entanglement, and Chaos: Exact Results' where this work was completed.


\end{document}